\documentclass{emulateapj}

\def\etal{et al.\ }
\def\eg{{\it e.g.\ }}

\def\p3m{P${}^3$M}
\def\ap3m{AP${}^3$M}
\def\-{{\em{---}}}
\def\msun{{M_\odot}}
\def\E3{{\cal E}_{\rm g}^{III}}

\newcommand{\be}{\begin{equation}}
\newcommand{\ba}{\begin{eqnarray}}
\newcommand{\ee}{\end{equation}}
\newcommand{\ea}{\end{eqnarray}}

\begin{document}
\title{The Spatial Distribution of the Galactic First Stars I: 
  High-Resolution N-body Approach}
\author{Evan Scannapieco$^{1}$, Daisuke Kawata$^{2,3}$, Chris B. Brook$^{4}$, 
Raffaella Schneider$^{5}$,\\ Andrea Ferrara$^{6}$, and Brad K. Gibson$^{7}$}
\altaffiltext{1}{Kavli
Institute for Theoretical Physics, Kohn Hall, UC Santa Barbara, Santa
Barbara, CA 93106} 
\altaffiltext{2}{The Observatories of the Carnegie Institution of Washington, 
813 Santa Barbara St., Pasadena, CA 91101}
\altaffiltext{3}{Swinburne University of Technology, Hawthorn VIC 3122, Australia}
\altaffiltext{4}{D\'epartement de physique, de g\'enie physique et d'optique,
Universit\'e Laval, Qu\'ebec, QC, Canada  G1K 7P4}
\altaffiltext{5}{INAF/Osservatorio Astrofisico di Arcetri, Largo Enrico Fermi 5, 50125 Firenze, Italy}
\altaffiltext{6}{SISSA/International School for Advanced Studies, Via
Beirut 4, I-34914, Trieste, Italy.}
\altaffiltext{7}{Centre for Astrophysics, University of Central Lancashire, 
  Preston, PR1 2HE, United Kingdom}
\begin{abstract}

We study the spatial distribution of Galactic metal-free stars by
combining an extremely high-resolution ($7.8 \times 10^{5} \msun$ per
particle) Cold Dark Matter N-body simulation of the Milky-Way with a
semi-analytic model of metal enrichment.  This approach allows us to
resolve halos with virial temperatures down to the $10^4$K atomic
cooling limit, and it is sufficiently flexible to make a number of
robust conclusions, despite the extremely uncertain properties of the
first stars.  Galactic metal-free stars are formed over a large
redshift range,  which peaks at $z \approx 10$, but continues down to
$z \approx 5,$ contributing stars at wide range of Galactocentric
radii.  Stars containing only metals generated by primordial stars are
similarly  widespread. Neither changing the efficiency of metal
dispersal by two orders of magnitude, nor drastically changing the
approximations in our semi-analytical model can affect these result.
Thus, if they have sufficiently long lifetimes, a significant number
of stars formed in initially primordial star clusters should be found
in the nearby Galactic halo regardless of the specifics of metal-free
star formation.  Observations of metal abundances in Galactic halo stars
should be taken as  directly constraining the properties of primordial
stars, and  the lack of metal-free halo stars today should be taken as
strongly suggesting a $0.8 \msun$ lower limit on the primordial
initial mass function.

\end{abstract}


\section{Introduction}

Metals are made in stars, but stars need not to have metals.
While the early universe was an efficient source of helium,  
primordial nucleosynthesis was halted before it was able to produce
elements heavier than lithium.  Thus the first stellar generation of
Population III (PopIII) stars condensed from a gas that was free of
carbon, silicon, iron, or any of the other heavy elements that are 
prominent today.

Indeed, to date not a single star has been observed that does not
contain metals.  Instead stellar nucleosynthetic products are
everywhere.  The lowest metallicity galaxies known are enriched to
$\sim 0.02Z_\odot$ (Searle \& Sargent 1972); the
lowest-density regions of the intergalactic medium (IGM) appear to be
enriched out to the highest redshifts probed  (Schaye \etal 2003;
Pettini \etal 2003; Aracil \etal 2004); and extensive searches for
metal-poor halo stars have failed to uncover any PopIII candidates
beyond a handful of stars  with extremely low abundances of elements heavier
than magnesium (Christlieb \etal 2002; Cayrel \etal 2004).

Interpretation of these stellar abundances is especially complicated
by the fact that searches for metal-poor stars are targeted at the
Galactic halo, where dust extinction is minimal and crowding is not a
serious concern.   These analyses have provided a number intriguing
constraints on  the enrichment history of the halo (Freeman \& Bland-Hawthorn 2002;
Beers \& Christlieb 2005),  including detailed measurements of the
metallicity distribution function from [Fe/H] $\sim -5$ to [Fe/H]
$\sim -1$ (\eg Ryan \& Norris 1991; Beers \etal 1992; Barklem \etal
2005), a detection of a shift in iron peak ratios below [Fe/H] $\sim
-3$ (McWilliam \etal 1995; Carretta \etal 2002; Cayrel \etal 2004,
Cohen \etal 2004), and the presence of extremely heavy element
deficient stars in
which [C/Fe] $> 2.0$ (Christlieb \etal 2002; Frebel \etal 2005).  While
these and other observations have provided the basis for several
theoretical analyses of the first stars (\eg Hernandez \& Ferrara 2001;
Scannapieco, \& Broadhurst 2001; Oey 2002; Schneider \etal
2003;  Tumlinson 2006), these are subject to substantial uncertainties 
due to the unknown spatial distribution of such objects.

In fact, it is still unclear how the observed population of halo
stars is related to PopIII star formation.  Indeed, White \& Springel 
(2000) used high-resolution cosmological simulations to show that the oldest
stars are naturally strongly concentrated towards the center of
the Galaxy, but stars that form in the full population of 
$T_{\rm vir} \approx 10^4$ K halos are widely distributed.
More recently, Diemand, Madau, \& Moore (2005)
used a resimulation technique to track the positions of particles
contained in the protogalactic halos collapsing from very rare 
$3.0 \sigma$ and $3.5 \sigma$ perturbations (where $\sigma^2$ is 
the variance of linear fluctuations on the scale of the objects).
Associating PopIII stars with
$3.0 \sigma$ perturbations, they found that the density of such
objects in the Galactic halo at the radius of the solar orbit 
is three orders of magnitude
lower than in the bulge.   Associating PopIII stars with $3.5 \sigma$
perturbations led to even more extreme results, decreasing the
number of first stars at the solar orbital radius by almost another order 
of magnitude.  Similarly, Karlsson (2006) was able to reproduce
a ``metallicity desert'' between [Fe/H] = -5 and -4 by associating
metal-free star formation with only the very earliest forming
stars. 

Yet this association of metal-free stars with the highest density regions
is likely to be misleading.  As shown in  Scannapieco, Schneider, \&
Ferrara (2003, hereafter SSF03; see also Schneider \etal 2006) 
cosmological enrichment was a local process, which occurred in 
different regions over an extended redshift range.  
Thus while the peak of PopIII star formation is likely to
have occurred at $z \approx 10$, such stars should  have continued to
form at  much lower redshifts. Furthermore, in these models metal free
stars  do not form in perturbations of a given $\sigma$, but rather
within $10^{7.5} \msun - 10^{8} \msun$ objects, which are just large
enough to cool within a Hubble time, but small enough that  they  are
not clustered near areas of previous star formation.

This spread in $\sigma$ and formation redshift raises the prospect
that the remnants of metal-free protogalactic clouds would have indeed
ended up in the halo.  This would mean that the absence of observed
metal-free objects might require the first stars to form with an
initial mass function (IMF) biased to stars with
lifetimes shorter than  a
Hubble time, and  perhaps even biased to masses $\ge 100 \msun$ as
suggested by a number of recent theoretical studies (\eg  Nakamura \&
Umemura 1998; Abel, Bryan, \& Norman 2000;  Bromm \etal 2001;
Schneider \etal 2002; Ripamonti \etal 2002; Tan \& McKee 2004).
Furthermore, it would mean that some of the stars with unusual yields
found in the halo of our galaxy are indeed providing us with direct
information about the yields  of PopIII stars.

In this paper we combine an extremely high-resolution N-body
simulation  of the formation of the Milky-Way with a semi-analytical
model of metal enrichment that closely parallels that discussed in
SSF03.  While we plan to conduct a careful comparison of this approach
with a more direct chemodynamical model of the Milky-Way in a
companion publication (Brook et al.\ in prep.), 
the dark-matter based model described here
nevertheless has a number of distinct advantages.  Although gas
cooling is essential to reproduce the spatial distribution of  late-forming
metal-rich Galactic disk stars, the spatial distribution of 
low-metallicity stars in the bulge and halo can be approximated 
from an N-body point of view. Such
stars were formed at early times in star clusters contained  in much
smaller dark matter halos, and they subsequently  evolved as
collisionless particles. Thus by carefully tagging N-body  particles
directly in progenitor halos as star-particles, we can use  them to infer
the positions of the earliest stellar populations.  This approach
allows us to work at a mass resolution $(7.8 \times 10^5 \msun$ per
dark-matter particle) that is currently impossible in gas simulations
of the Milky-Way, which means that we can capture the formation of all
halos within which efficient star-formation is expected to occur.

Our semi-analytic model of metal enrichment, on the other hand, gives
us the flexibility to use this simulation to study a wide range of
models for the properties of the first stars.  As the typical mass of
such stars remains uncertain within at least two orders of magnitude,
the metal and kinetic energy output from SNe arising from PopIII stars
remains completely unknown.  This uncertainty is further compounded by
the possibility that metal-free stars in the $\approx 200 \msun$ mass
range may have given rise to tremendously powerful pair-production
supernovae (\eg Heger \& Woosley 2002), with kinetic energies up to
$10^{53}$ ergs per event.   While exploring this full range of
possibilities by direct simulations is extremely computationally
expensive,  our technique allows us to carry out this parameter
study in a few minutes, leading to a number of robust conclusions
despite the large uncertainties involved.

The structure of this work is as follows.  In \S 2 we describe our
simulations and semi-analytical model.  In \S 3 we apply our model to
consider the distribution of the first and second stars in a wide range
of cosmological scenarios.   Conclusions are given in \S 4.

\section{Method}

To identify the location of PopIII stars in our galaxy, we combine 
a high-resolution N-body simulation of the formation 
of a Milky-Way 
size galaxy with  a semi-analytical model of metal enrichment.
In \S2.1 we describe our simulation in detail,
as well as the method we use to extract gravitationally bound
halos.  In section \S2.2, we describe our semi-analytical model and
discuss how it is used to construct the expected positions
for stars forming from gas that is initially primordial, which we label
as ``the first stars,'' and stars forming from gas that is initially 
enriched by only primordial stars, which we label as ``the second stars.''

\subsection{Simulation and Identification of Halos}

Throughout our study  we assume a Cold Dark Matter model with parameters $h=0.71$,
$\Omega_0 h^2$ = 0.135, $\Omega_\Lambda = 1-\Omega_0$, 
$\Omega_b h^{2} = 0.0224$, $\sigma_8 = 0.9$, and $n=1$, 
where $h$ is the Hubble constant in units of 100 km/s/Mpc,
 $\Omega_0$, $\Omega_\Lambda$, and $\Omega_b$
are the total matter, vacuum, and baryonic densities in units of the
critical density, $\sigma_8^2$ is the variance of linear fluctuations
on the $8 h^{-1}{\rm Mpc}$ scale, and $n$ is the ``tilt'' of the
primordial power spectrum (\eg Spergel \etal 2003). 
Our N-body simulation is carried out with the {\tt GCD+} code
(Kawata \& Gibson 2003a), and it uses a multi-resolution technique 
(Kawata \& Gibson 2003b) to achieve high-resolution in the regions of
interest, while the outer regions exerting tidal forces 
are handled with lower resolution.

The initial conditions for these simulations at $z=56$ are constructed
using the public software {\tt GRAFIC2} (Bertschinger 2001).  The
low-resolution region is a $20h^{-1}$ Mpc diameter sphere selected
from a low-resolution cosmological simulation, for which an isolated
boundary condition is applied.  The highest resolution region is a
sphere with a radius four times the virial radius of the system at
at $z=0$.  In this region, the Dark Matter (DM) particle mass and softening length
of are $M_{\rm vir}=7.8\times10^5$ ${\rm M}_\odot$  and $r_{\rm
vir}=540$ pc, respectively.  The system consists about $10^6$
particles within the virial radius. The virial mass and radius of the
system at $z=0$ are $M_{\rm vir}=7.7\times10^{11}$ ${\rm M}_\odot$ and
$r_{\rm vir}=239$ kpc, which are roughly consistent with the estimated
mass ($M_{\rm vir}=1.0\times10^{12}$ ${\rm M}_\odot$) and 
virial radius ($r_{\rm vir}=258$ kpc) of the Milky Way (Klypin, Zhao,
\& Somerville 2002; Battaglia \etal 2005,  see also Dehnen,
McLaughlin, \& Sachania 2006 for a slightly more massive, $M_{\rm
vir}=1.5\times10^{12}$ ${\rm M}_\odot$, and compact, $r_{\rm vir}=200$
kpc, model.)  The initial conditions are essentially the same as the
KGCD model in Bailin \etal (2005), but  the current simulation has a
much higher resolution and slightly different cosmological parameters.

Using a low-resolution simulation including gas physics and star
formation, we confirm that this initial condition will lead to a disk
galaxy.   A closer comparison between our results here  and a
lower-resolution gas-dynamical model 
will be presented in a future publication (Brook \etal in prep.).  
While gas physics is
essential to reproduce the spatial distribution of disk stars, an
N-body approach is appropriate for the halo and bulge populations on
which we are focused here.  Such stars should initially form in much
smaller gas disks condensing in neighboring halos, and later interact
as  collisionless particles.  Thus by carefully tagging N-body
particles directly in progenitor halos as star-particles, we can use
them to infer the positions of the first and second stars in our final
galaxy.

The simulation data is output every $0.11$ Gyr, and at each output, we
use a friend-of-friends group finder (FOF, Davis \etal 1985) to
identify  the virialized DM halos. We apply a standard approach with a
linking parameter of $b = 0.2$ and threshold number of particles of
50, and we explore the results of varying this parameter below.
We then construct a merger tree, associating the halos at
different time steps.  In particular, we search for the halo at
each time step $n$ that contains the largest number of the member
particles  of a given halo at the previous time step, $n-1.$
We define the virial mass and radius for each halo, taking into account
the cosmology and redshift, following the fitting formula in the
Appendix of Kitayama \& Suto (1996).

A key issue is the minimum mass of the halos in which  stars formed.
Before reionization, this depends sensitively on the presence of $H_2$
(\eg O'Shea \etal 2005).  In small objects, molecular hydrogen is
easily photodissociated by  11.2-13.6 eV photons, to which the
universe is otherwise transparent.   Thus  emission from the very
first stars quickly destroyed all avenues for $H_2$ cooling (Dekel \&
Rees 1987; Haiman, Rees, \& Loeb 1997; Ciardi, Ferrara, \& Abel 2000).
This raised the minimum virial temperature necessary to cool
effectively to approximately $10^4$ K, although the precise value of
this transition is the subject of debate  (\eg Glover \& Brand 2001;
Yoshida \etal 2003) and is somewhat dependent on the level of the
high-redshift X-ray background (Haiman, Rees, \& Loeb  1996; Oh 2001;
Machacek, Bryan, \& Abel 2003).

A natural question is whether the few stars that dissociated
primordial molecular hydrogen may have been also enriched
a significant fraction of the intergalactic medium.  To
estimate this fraction we consider two cases, one in which
the primordial stars that contributed most to the dissociating
background were  $\approx 10 \msun$, and a second case in which
the primordial IMF was peaked to form very massive stars $\approx 
200 \msun$.   In the $10 \msun$ case the rate of dissociating
photons is $\approx 10^{47}$ s$^{-1}$ (Glover \& Brand 2001; see also
Schaerer 2002; Venkatesan, Tumlinson, \& Shull 2003) and the
stars live $30$ Myrs (Fagotto \etal 1994) to generate
a total of $\approx 10^{62}$ dissociating photons per star.  In
the 200 $\msun$ case the dissociating rate is 
$\approx 2 \times 10^{49}$ s$^{-1}$, but the lifetime is only 
$2$ Myrs, resulting in a total of $\approx 10^{63}$ such photons 
per star.  Combining these values with the $3 \times 10^{-4}$
mean cosmological fraction of $H_2$ (\eg Yoshida \etal 2003), we can
estimate the typical radius of a region dissociated by a primordial
star as $r_{\rm diss} = 200 \Delta^{-1/3}$ comoving kpc 
in the 10 $\msun$ case and 
$r_{\rm diss}  = 450 \Delta^{-1/3}$ comoving kpc in the 
200 $\msun$ case, where 
$\Delta = \rho/\bar \rho$ is the mean overdensity of the 
dissociated region.

On the other hand, adopting a simple Sedov-Taylor estimate
of the maximum distance that a given supernova will be able to
distribute metals as a function of time gives 
$r_{\rm enrich} 
= 25 (1+z)^{2/5} t_{\rm Gyr}^{2/5} E_{\rm 51}^{1/5} \Delta^{-1/5}$ comoving kpc
where $t_{\rm Gyr}$ is the expansion time in Gyrs and $E_{\rm 51}$ is the
energy input in units of $10^{51}$ ergs.  If we assume no kinetic
energy losses, replace $t_{\rm Gyr}$ with the Hubble time, and 
take a typical redshift of 15, this gives 
$r_{\rm enrich} = 35 E_{\rm 51}^{1/5} \Delta^{-1/5}$ comoving kpc.  

Since the formation of these very first stars is self-truncated by
$H_2$ dissociation, they should cease to form and distribute
metals when the filling factor of their dissociation regions reaches
unity. This means that the relative ratio of the dissociation/metal enrichment
volume should be an accurate predictor of the total filling factor of 
such very early metals.  In the 10 $\msun$ case  this ratio is
$(r_{\rm enrich}/r_{\rm diss})^{3} \approx 0.005$
while in the $200 \msun$ case
$(r_{\rm enrich}/r_{\rm diss})^{3} \approx 0.002$
where we have taken a typical SN energy of $10^{52}$ ergs
(Heger \& Woosley 2002). 
Thus less than 1\% of the universe is likely to have been
enriched by this earliest generation of stars, and this
only in the very densest regions.
Rather the majority of primordial enrichment  is should have occurred
in larger objects with $T_{\rm vir} \geq 10^4$ K. 

 In this case,
(sometimes referred to Population II.5; \eg Grief \& Bromm 2006)
efficient atomic-line cooling establishes a dense locally-stable disk,
within which non-equilibrium free electrons catalyze $H_2$. Unlike in
less massive halos, $H_2$ formation in these objects is largely
impervious to feedback from external UV fields, due to the  high
densities achieved by atomic cooling (Oh \& Haiman 2002).  As
discussed in SSF03, regions of late primordial star formation  may
have already been detected as a subgroup of high-redshift Lyman-alpha
emitters, although distinguishing them from Population II/I stars is
extremely difficult (Dawson \etal 2004).

In this study then, we adopt a fixed threshold of  $T_{\rm vir} \geq
10^4$ K for star formation in halos at all redshifts.  In our
cosmology, and for the high redshifts relevant for the formation of
primordial objects,  this corresponds to a minimum mass of 
\be 
M_{\rm min} \equiv 3.0 \times 10^9 (1+z)^{-3/2} M_\odot.  
\label{eq:mmin}
\ee 
While at
high-redshift this mass can be over four orders of magnitude smaller
than the final virial mass of the Galaxy, our high-resolution simulation
nevertheless makes it possible to trace these objects.  This means 
that at redshifts below 17,  all the halos relevant to our study will 
contain at least 50 particles and be well-identified by our FOF group finder.
However, at the very earliest redshifts $\geq 17$ such halos will only
be marginally resolved.

\subsection{Identification of First and Second Objects}

Having constructed the history of  dark matter halos as a function of
redshift,  our next step is to use this to identify two types of
objects: i) halos that collapse out of primordial gas, which we
identify as Population III objects containing ``the first stars,'' and
ii)  halos that collapse from gas that has been enriched purely by
material ejected from the Population III objects, which we identify as
second-generation objects containing  ``the second stars.''  While we
vary the IMF of the first stars over a wide range of possibilities, we
assume that the second stars  form with an IMF similar to that seen
today, although they are likely to display peculiar enrichment
patterns.

It is important to make clear that our model is not able to say anything
about the progression of star formation and self-enrichment
within a particular halo (see \eg Susa \& Umemura 2006).  
Rather, {\em our labels purely reflect the properties of the most
metal-poor stars formed in any given region.}
Thus it is entirely possible that some fraction of
the stars forming in halos that collapse from  primordial gas will
nevertheless contain metals that were inherited from other stars
forming in the same temporally-extended burst.   For our purposes here
then, stars formed in such self-enriched primordial star clusters are
included in the distribution of  ``the first stars.''  
Similarly  we exclude all halos that are direct descendents of first objects 
from our list of second generation objects, as these would exist cospatially 
with stars that were truly primordial.

These definitions are intended to extract the most information possible from
our modeling approach.  For observers, anywhere labeled first stars
would be a good place to search for metal-free stars, if they are sufficiency 
long-lived to survive until today.   Similarly, both regions
labeled first stars and second stars in out method would be a good places
to search for stars enriched purely by the products of metal-free stars.
From a theoretical point of view, the spatial distribution of first stars in our models
should be directly comparable to the spatial distribution of 
metal-free stars in future detailed gas simulation, while the spatial distribution of
{\em both} first and second stars should be comparable to the spatial distribution
of PopIII-enriched stars in such simulations.
These definitions are important to keep in mind when interpreting our results, 
as emphasized below.

As the distribution of both these populations of stars is primarily dependent on the
highly-uncertain properties of Population  III objects, our goal is
 to develop a simple model of metal dispersal that allows us explore a
 large range of possibilities in a straightforward way.   Following
 our approach in SSF03, we adopt a model of outflows as spherical
 shells expanding into the Hubble flow (Ostriker \& McKee 1988;
 Tegmark, Silk \& Evrard 1993), for both PopIII and PopII/I objects.
 These shells are assumed to be driven only by their internal hot gas
 pressure and decelerated by inertia due to accreting material and
 gravitational drag  while escaping from the host.  
In this case there are two relevant evolutionary equations: one 
that describes
the change in the velocity of the bubble, 
\be 
\ddot{R_s} = \frac{3
 P_b}{\bar \rho R_s} - \frac{3}{R_s}(\dot{R_s} - HR_s)^2 - \Omega_M
 \frac{H^2 R_s}{2},
\ee
where the terms on the right hand side correspond to internal pressure,
accumulation of material from the surrounding medium, and gravitational
drag, respectively, 
and a second equation that describes the evolution of the internal
energy of the bubble,
\be
\dot{E_b} =  L(t) - 4 \pi R_s^2
 \dot{R_s} P_b,  
\ee 
where the terms on the right hand side correspond to
energy input from the supernovae and the $P dV$ work done by the interior gas.
In these equations the overdots represent time derivatives,
the subscripts {\it s} and {\it b} indicate shell and bubble
quantities respectively, $R_s$ is physical radius of the shell, $E_b$
is the internal energy of the hot bubble gas, $P_b$ is the pressure
of this gas, and $\bar \rho$ is the mean IGM background density.
Here we assume adiabatic expansion with an index $\gamma=5/3$ such
that $P_b=E_b/2\pi R_s^3.$

The evolution of each such bubble is completely determined by the
mechanical luminosity evolution assigned, $L(t)$.  Again for
simplicity, we take both PopIII and PopII/I objects to undergo a
starburst phase of $t_{\rm SN} = 5 \times 10^7$ years, but with
different prescriptions for the total energy input into the wind.  
We approximately account for the gravitational potential of the host
galaxy by subtracting the  value of $G M^2 (\Omega_b/\Omega_M) /r_{\rm
vir}$ from the total wind energy, where $M$ is the total mass (dark +
baryonic) of the object and $r_{\rm vir}$ is its virial radius.
Approximating the collapse overdensity by a constant 
value of 178, the overdensity associated with a virialized cosmological
object, then gives a mechanical luminosity of 
\ba 
\label{eq:meclum}
L(t) & = & 160 L_\odot  \left[
f_\star^{III,II}  f_w  E_{\rm kin}^{III,II} {\cal N}^{III,II}  \right. \\ \nonumber
& & \left. - 5 \times 10^{-12} (M_b h \Omega_M/\Omega_b  )^{2/3} (1+z_c) \right] M_b
\Theta(t_{\rm SN}-t),
\ea  
where $f_\star$ is the fraction of gas converted into stars,
$f_w$ is the fraction of the SN kinetic energy that is channeled into
the galaxy outflow, $E_{\rm kin}$ in the kinetic energy per supernova,
${\cal N}$ is the number of supernovae per $\msun$ of stars, $M_b$ is
the baryonic mass of the galaxy in units of solar mass, $z_c$ is
the collapse redshift of the object, and $\Theta$ is the Heaviside
step function. Note the precise value of $t_{\rm SN}$ is not
important as long as it short relative to the $\approx 500$ Myr 
Hubble time at the redshift in which we are most interested.  However,
the value of the prefactor in front of $E_{\rm kin}$ in eq.\ (\ref{eq:meclum}) 
is defined by this choice.   Halving $t_{\rm SN}$, for example
would require for us to double this factor but have only a very
weak effect on the time that the bubbles reach their neighbors,
and hence on the final distribution of first and second stars.
Note also that the gravitational
drag is only important in objects with total masses $\gtrsim 10^{11}
(1+z)^{-3/2} \msun,$ and thus has only a minor impact on our results
here.

The only difference in the wind evolution of PopIII and PopII/I arises
from the product, $f_\star^{III,II} f_w E_{\rm kin}^{III,II} {\cal
N}^{III,II}$, which, again  following SSF03, we define as the ``energy
input per unit gas mass''  ${\cal E}_{\rm g}^{III,II}$, which we
express below in units of ${10^{51} {\rm erg} \, M_\odot^{-1}}$.  To
determine this value in the PopII/I case, we take $f_\star^{II}$ to be
$0.1$, which gives good agreement with the observed high-redshift star
formation rates and abundances of metals measured in 
quasar absorption line systems (Thacker, Scannapieco, \& Davis 2002;
Scannapieco, Ferrara, \& Madau 2002).  Also in this case, we constrain
$f_w$ by combining the overall efficiency of 30\% derived for the
$2\times 10^8 M_\odot$ object simulated by Mori, Ferrara, \& Madau
(2002) with the mass scaling derived in Ferrara, Pettini \& Shchekinov
(2000), which was obtained by determining the fraction of starburst
sites that can produce a blow-out in a galaxy of a given mass.  This
gives $f_w(M) = 0.3\delta_B(M)/\delta_B(M=2\times 10^8 M_\odot)$ where
\be 
\delta_B(M)=
\cases{ 
1.0 & $\tilde N_t \leq 1$ \cr 
1.0 - 0.165 \, {\rm ln} (\tilde N_t) &
 $1 \leq \tilde N_t \leq 100$ \cr 
[1.0 - 0.165 \, {\rm ln} (100)] \, 100 \, \tilde N_t^{-1}  & 
$ 100 \leq \tilde N_t$ \cr},
\label{eq:db}
\ee 
and $\tilde N_t \equiv 1.7 \times 10^{-7} (\Omega_b/\Omega_M) M/M_\odot$ is 
a dimensionless parameter that increases linearly with the overall number 
of SNe produced in a starburst, divided by the star formation efficiency, 
$f_\star^{II}.$

In the PopIII case, on the other hand, there are no direct constraints
on either $f_\star^{III}$ or the wind efficiency.  For these objects
we allow these parameters to be free, varying $\E3$ over a large range
as discussed below.  Finally, when outflows slow down to the point
that they are no longer supersonic, our approximations break down, and
the shell is likely to be fragmented by random motions. At this point we
let the bubble expand with the Hubble flow.  These prescriptions
then give us a complete model of $R_{III,II}(M, z,z'),$ the
comoving radius of an outflow at a redshift $z'$ associated with a 
PopIII or PopII/I halo of mass $M$ collapsing at a redshift $z.$

Our goal is then to tag halos that form outside of all expanding
shells as first stars objects, and to tag halos that form within
PopIII outflows as second-stars objects. 
In our fiducial model we assume that in each such halo stars 
form with the same radial profile as the dark matter, although we also vary 
this  assumption in \S2.2.
As we are concerned only with whether a halo is contained
within {\em any} such wind from a neighbor or direct progenitor, we
need not worry about the possibility  of double counting by
identifying the same collapsed object at multiple outputs.   Rather to
determine if a halo forming at a redshift $z'$ has been enriched, we
can safely compute $R_{III,II}(M, z,z')$ for each and every object
forming with $z > z'$ and determine if any of them has overtaken the
lower-redshift object.   In this case the impact of a given halo on its
neighbors as well as its direct descendants
will be  determined by the maximum $R_{III,II}[M(z), z,z']$
computed at any $z > z'$, while all other smaller shells will have no
impact on the calculation.

While this picture neglects the precise density and infall structure
around each forming object, we are able to quantify the errors arising
from this approximation by adopting two methods for computing the
distance between two objects. Note however that  as the
outflows we are most concerned with occur at very high redshift, the
nonlinear evolution of structure will not be as severe an issue as it
is for lower-redshift winds.

One method of computing distances, which we shall refer to as
``Eulerian,'' allows  each wind to move along with the center of mass
of its central halo.  This means that as we move between each output
we compute the most likely daughter halo associated with each higher
redshift outflow and re-center it around the position of this object.
This takes advantage of the nonlinear motions  computed in the
simulation, but is likely to overestimate the impact  of winds as it
does not take into account the impact of infall around collapsing
structures.

A second method of computing distances, which we shall refer to  as
``Lagrangian,''  works completely in the reference frame in which the
position of each halo is computed as the center of the mass of the
{\em original} $z=55.7$ positions of the particles it contains.  While
this approach neglects much of the nonlinear information available
from the simulation, it is nevertheless likely to represent a more fair
comparison as: i.) our thin-shell solution, which assumes expansion in
the Hubble flow, is essentially already working in this coordinate
system;  ii.)  enrichment should be more closely dependent on
the column depth between the source and recipient halos than on their
physical separation.  Furthermore this approach most directly
parallels that of our previous analytical models (SSF03), and thus we
adopt it as our fiducial  method in the results presented below.

\section{Results}

In this section we summarize our results for a wide range of parameters choices
and assumptions, the details of which is described in detail below.   For reference
purposes, the specifics of each of these runs is summarized in Table 1.

\begin{table}
\begin{center}
\caption{\footnotesize Properties of First Stars Models}
\begin{tabular}{|l|lllll|} \hline
Model &      &  Distance & Linking & Distribution & Minimum \\
Name    &  $\E3$ &  Measure  & Length  & of Stars   &   $T_{\rm vir}$\\
\hline
2L   & $10^{-2}$  & Lagrangian & 0.2 & Follows DM & $10^4$ K \\
3L   & $10^{-3}$  & Lagrangian & 0.2 & Follows DM & $10^4$ K \\
4L   & $10^{-4}$  & Lagrangian & 0.2 & Follows DM & $10^4$ K \\
2E   & $10^{-2}$  & Eulerian & 0.2     & Follows DM & $10^4$ K \\
3E   & $10^{-3}$  & Eulerian & 0.2      & Follows DM & $10^4$ K \\
4E   & $10^{-4}$  & Eulerian & 0.2      & Follows DM & $10^4$ K \\
2Lc   & $10^{-2}$  & Lagrangian & 0.2   & Central 10\% & $10^4$ K \\
3Lc   & $10^{-3}$  & Lagrangian & 0.2   & Central 10\% & $10^4$ K \\
4Lc   & $10^{-4}$  & Lagrangian & 0.2   & Central 10\% & $10^4$ K \\
2L15   & $10^{-2}$  & Lagrangian & 0.15 & Follows DM & $10^4$ K \\
3L15   & $10^{-3}$  & Lagrangian & 0.15 & Follows DM & $10^4$ K \\
4L15   & $10^{-4}$  & Lagrangian & 0.15 & Follows DM & $10^4$ K \\
2LRc   & $10^{-2}$  & Lagrangian & 0.2 & Central 10\% & $10^4$ K/$5\times10^4$ K \\
3LRc   & $10^{-3}$ & Lagrangian & 0.2 & Central 10\% & $10^4$ K/$5\times10^4$ K \\
4LRc   & $10^{-4}$  & Lagrangian & 0.2 & Central 10\% & $10^4$ K/$5\times10^4$ K \\
\hline
\end{tabular}
\end{center}
\end{table}

\begin{figure}
\plotone{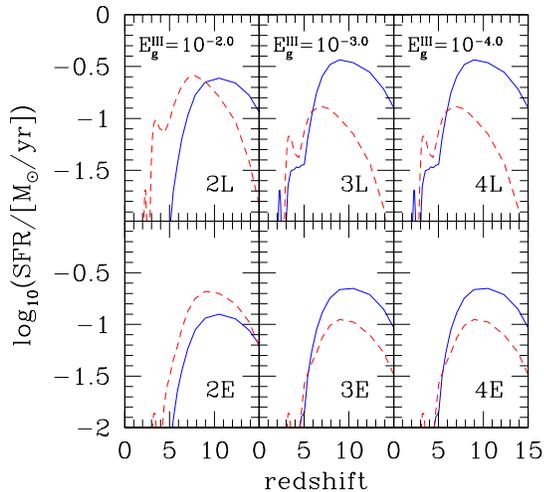}
\caption{Star formation rate as a function of redshift in our models.
In each panel, the solid lines show the rate in first stars objects
containing initially primordial gas, while the dashed lines show the
star-formation rate in second-stars objects, which have been
pre-enriched by only the products of primordial stars.  From left to
right the panels represent models ranging  from those containing
extremely strong ($\E3 = 10^{-2}$) winds, to those with much more
gradual ($\E3 = 10^{-4}$) metal ejection. A 10\% star formation
efficiency is assumed in all cases.  The top row shows the results of
models in which the distances between halos have been computed using
the ``Lagrangian method'' described in  \S2.2, and the bottom row
represents models in which distances have been computed using the
``Eulerian method,'' which predicts a somewhat stronger impact from
winds.  Each panel is labeled by its corresponding model name as
described in Table 1.  Despite the large range of parameters and
methods considered, metal-free star formation occurs down to $z
\approx 3$ in many cases, and down to $z \approx 5$ even in the most
extreme estimates.}
\label{fig:SFR}
\end{figure}

\begin{figure*}
\epsscale{0.9}
\plotone{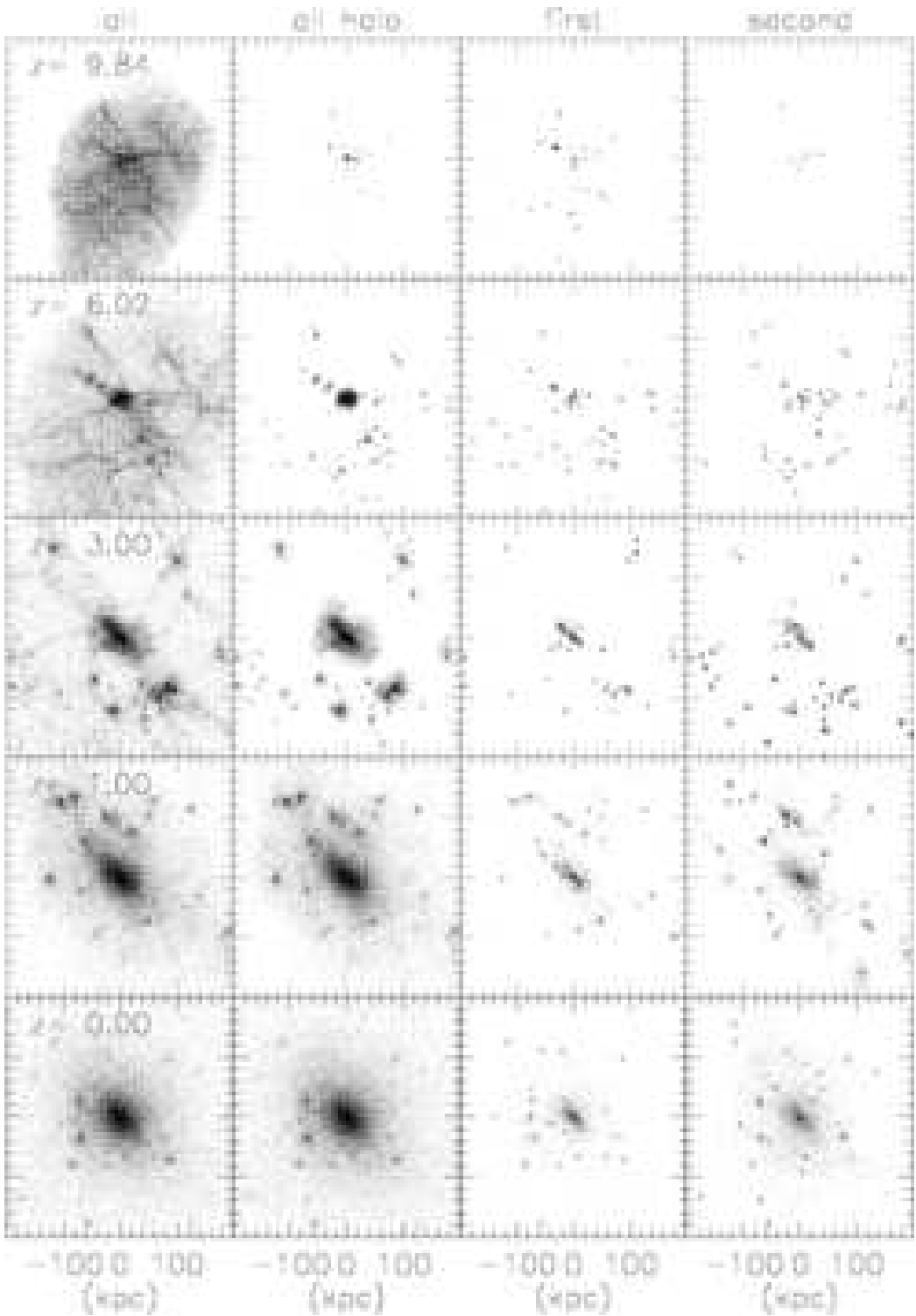}
\caption{Distributions of first and second stars  (3rd and 4th columns) 
in a a 200 proper kpc$^3$ region, at various redshifts, in model 2L,
a Lagrangian model with $\E3 = 10^{-2}$.
For comparison, the 1st and 2nd columns show all the particles and
all the particles within halos with masses
above $M_{\min}$ in equation (1), respectively.
In the 1st and 2nd columns, to avoid confusion, 
we only plot 1/10 of the whole sample.}
\label{fig:anim}
\end{figure*}

\begin{figure*}
\epsscale{0.8}
\plotone{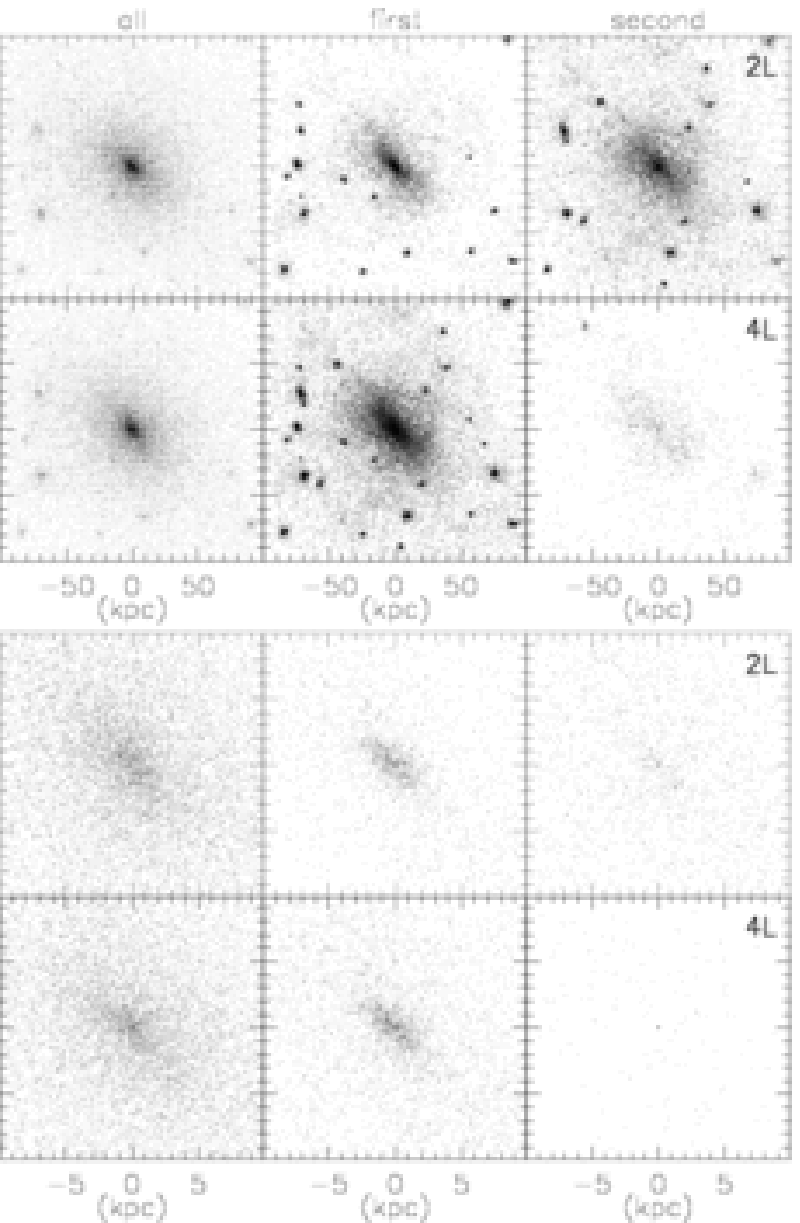}
\caption{Distributions of first and second stars at $z=0$ in
our fiducial, Lagrangian models.  
From top to bottom, the rows correspond to the central
100 kpc region in the $\E3 = 10^{-2}$ model,
the central 100 kpc in the $\E3 = 10^{-4}$ model,
the central 10 kpc in the $\E3 = 10^{-2}$ model, and
the central 10 kpc in the $\E3 = 10^{-4}$ model.
In interpreting this figure, it
is important to note that we are explicitly ignoring mass loss due
to stellar evolution.}
\label{fig:Dist1}
\end{figure*}

\subsection{Fiducial Modeling}

In Figure \ref{fig:SFR}, we show the star formation rate in  first and
second stars as a function of redshift,  for a wide range of feedback
efficiencies and for both Lagrangian and Eulerian distances,
corresponding to runs 2L-4L and 2E-4E in Table 1 respectively.  When
computing the star formation rate, we consider only stars that end up
gravitationally bound to our final galaxy, so as to exclude objects
formed near the boundary of the simulation volume.  In all cases we
assume $f_\star^{III} = f_\star^{II} = 0.1$ such that 10\% of the gas
mass in each PopIII halo is converted into stars.  Note that while
this value is consistent with the observed star formation rate of
normal stars at lower redshifts (\eg Scannapieco, Ferrara, \& Madau
02), it is certainly possible that metal-free star formation was less
efficient and that a lower value such as $f_\star^{III} = 0.01$ is
appropriate in the PopIII case.   While this  normalization will
remain a primary uncertainty in our investigation, it can be easily
accounted for by rescaling our final results, and does not affect the
overall trends on which we are primarily focused.

In SSF03, we considered the range of  physically possible $\E3$ values
in detail.  An upper bound is provided by the case in which the
initial mass function (IMF) of  primordial stars is biased to very
high masses $> 100 M_\odot.$   While much of the stars in this mass
range would have died forming black holes, progenitors with masses
between $140$ and $260 \msun$  end their lives in tremendously
powerful pair-production supernovae.  In these SNe, $e^+ e^-$ pair
creation softens the equation state at the end of central carbon
burning, leading to rapid collapse, followed by explosive burning (\eg
Barkat, Rakavy, \& Sack 1967; Ober, El Eid, \& Fricke 1983; Bond,
Arnett, Carr 1984; Heger \& Woosley 2002).  The more massive the star,
the higher the temperature at bounce and the heavier the elements that
are produced by nuclear fusion.  In all cases, the star is completely
disrupted and the ejection energies ($\approx 3-100 \times 10^{51}$
ergs) are enormous.  In this case, choosing a  narrow Gaussian IMF
centered at 200 $\msun$, taking a 10\% star formation efficiency, and
fixing $f_w \approx 0.5$ places a generous upper limit of  $\E3 =
10^{-2},$ corresponding to the strongest feedback model in Fig.\
\ref{fig:SFR}.  Note that while this extreme model is useful to set an
upper limit on feedback, widespread enrichment from stars with such an
IMF is inconsistent with the observed abundance patterns of metal poor
stars  owing to the absence of pair-instability yields in the data
(Umeda \& Nomoto 2003; Tumlinson, Venktesan, \& Shull 2004).

A lower limit on $\E3,$ on the other hand, is computed by assuming
that such stars form with a typical Salpeter IMF, $f^{III}_\star
\approx 0.1$ and  $f_w \approx 0.1,$ which gives $\E3 \approx
10^{-4}.$  Not only does this model take metal ejection from PopIII to
be somewhat less  efficient than from PopII/I objects, but as shown in
SSF03, even less efficient models of PopIII metal ejection are so weak
that a large primordial star formation rate is predicted to this day.
Thus the weakest, $\E3 = 10^{-4},$ model shown in Figure \ref{fig:SFR}
can be taken to represent  a conservative estimate of the minimum
efficiency with which metals from PopIII objects were mixed into their
surroundings.

This large range of efficiencies of PopIII objects is what forces us
to adopt a semi-analytical model of winds in the first place.  Yet, as
shown in Figure \ref{fig:SFR}, robust general conclusions can be drawn
despite these uncertainties.   Firstly, the primordial objects that
predated the Milky-Way were  formed over an extremely large range of
redshifts, mirroring the overall cosmological PopIII distribution (see
SSF03).   Furthermore, for all $\E3$ values, and in both the Eulerian
and Lagrangian models, appreciable PopIII star formation continues
down to a relatively low redshift of $z \approx 5.$   In fact,
excluding the most extreme models, PopIII formation continues at a
substantial rate down to $z \approx 3.$

While increasing the efficiency of feedback has the impact of moving
the PopIII star formation to higher redshifts,  the overall $z \approx
10$ peak and $\Delta z \approx 5$ width of the epoch of PopIII
formation are surprisingly constant among models.  Instead, the
distribution of second stars is much more sensitive to these
uncertainties, ranging from an extended burst in the $\E3 = 10^{-2}$
case, to a weak high-redshift burst in the $\E3 = 10^{-4}$ case.

Again, we stress that these second stars are the stars forming in
halos that have been initally pre-enriched by PopIII neighbors, and
that a significant population of Pop-III enriched stars are also
likely to exist in ``first  stars'' halos forming of initially
primordial gas.  Finally, comparing the Eulerian and Lagrangian
models, we find, as expected, that Eulerian distances lead to somewhat
more efficient metal dispersal, although this is likely to be somewhat
of an overestimate.  Nevertheless the differences between these two
models are relatively mild for all choices of $\E3,$ giving us further
confidence in the  approximations involved in our semi-analytic
approach.

Figure \ref{fig:anim} demonstrates the positions of the first and
second stars at different
redshifts in the Lagrangian model with $\E3 = 10^{-2}$.  Here and
below we compute the masses as they would be if there were no mass
loss due to stellar evolution.  Thus these plots should be rescaled by
the fraction of first and second stars surviving to a given
redshift, model-dependent  quantities that we do not address here.

At early times ($z=9.84$), the first stars are forming close to the
central density peak of the progenitor galaxy, due
to the higher density peaks in this region.   Second stars form in
the halo in the neighborhood of first stars,  because they condense
from gas that is enriched by the material from the explosions from these
objects. At a later time ($z=6.02$), new first stars are still
forming, but now on the outer regions of the progenitor galaxy, because they
are not yet affected by the bubbles from the first stars that formed
preferentially in the central region.  As seen in the star formation
rate plot, Figure \ref{fig:SFR}, the formation of the first and second
stars is complete around $z=3$, at which time the full region is
enriched with metals. Also at this redshift, first and second stars
start to be accreted into the assembling Galactic halo.   This
assembly is almost complete by $z=1$, and thus the distribution at
this redshift is similar to that at $z=0$,  although at $z=0$ the
stars are more smoothly distributed.

In Figure \ref{fig:Dist1} we show the $z=0$ spatial distribution of
the first and second stars in closer detail, contrasting
two extreme models with strong $\E3 =
10^{-2}$ and weak $\E3 = 10^{-4}$ feedback.    From this
figure, we see that for all choices of $\E3$, the first or second
stars are not confined to the center of the Galaxy.   Rather, this
plots shows a  widespread distribution that is punctuated by dense
concentrations of stars associated with satellite galaxies, as seen
most clearly in the 100 kpc panels.  Furthermore, comparing the
satellite distributions between the two models shows that as one moves
from the strongest to the weakest feedback models, a large number of
the satellite second-generation objects switch over to  the
first-generation population.  This is because these objects were only
enriched by primordial stars, thus  a large number of them form from 
primordial gas if the winds from PopIII stars are weakened.
The more isolated the satellite galaxy, the more likely it is to 
contain primordial stars even in strong feedback models.

Focusing on the central 10 kpc of our simulation, which is more
representative of the Galaxy itself, we see that the distribution of
first and second stars, while peaked towards the center,
nevertheless contains many stars at large radii.  Again
this is true even in the  most extreme feedback models, and it
represents the single most important finding of our investigation.
{\em  If they have sufficiently long lifetimes, a significant number
of stars formed in initially primordial star clusters should be found
in the Galactic halo, regardless of the specifics of PopIII star
formation.}

\begin{figure}
\plotone{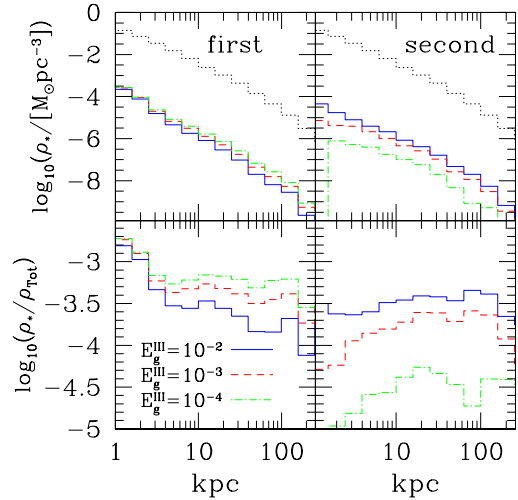}
\caption{Radial profile of first stars forming from initially primordial gas
(left column) and second stars formed from initially PopIII enriched gas
(right column) at $z=0$ in our fiducial Lagrangian models.  In the top
row, the dotted lines give the overall dark matter profile of the
galaxy, which is compared with the radial density of first and second
stars in  models 2L (solid), 3L (short-dashed), and 4L (dot-dashed).
The lower row shows the fraction of the total density
in first and second stars, with symbols as in the upper panels.
Like Fig.\ \ref{fig:Dist1} and those below, this plot does not
include any mass loss due to stellar evolution.
Furthermore, it assumes $f_*^{III}= f_*^{II} = 0.1.$}
\label{fig:radial1}
\end{figure}

To quantify this statement further, in Figure \ref{fig:radial1} we
plot the radial mass density of first and second stars, as compared to
the dark matter distribution in our simulation.  Again, in computing
these masses, we assume that 10\% of the gas mass  in each halo is
converted into stars  and we intentionally make no attempt to
account for mass loss due to the evolution of these stars.
Furthermore, we plot only out to 250 kpc, which is 1/4 the size 
of the high-resolution region.

Here we see that the density profiles of first and second stars are similar
to the total dark matter density profile, although the second stars
have a slightly shallower slope. As a result, the 
density of first stars at the center is 100 times higher
than at the 8 kpc orbital radius of the sun.
However, the important number for developing
observational strategies is the relative density of such stars
with respect to field stars.
While this is not directly computed in our simulation,
the lower panels of 
Figure \ref{fig:radial1} show the 
local mass density of stars normalized by the local 
density of dark matter.

Amazingly, the mass fraction contained  in the
first stars varies only very weakly with radius.  Moving from 1 to 100
kpc in the  $\E3 = 10^{-4}$ model, for example,  the fraction of the
mass in first stars decreases only by a factor $\approx 4.$
Furthermore, increasing the efficiency of PopIII winds to $\E3 =
10^{-2}$ has the effect of decreasing the fraction of first stars
without strongly affecting their radial distribution.
Note that in the inner region, the overall stellar density is expected
to be higher than dark matter density (e.g. Widrow \& Dubinski 2005),
that is the density of spheroidal stellar component of the Milky Way 
decreases with the radius more quickly than the dark matter
component. Therefore, the fraction of the first stars
compared with the field stars should be somewhat higher at large
radii than what Figure \ref{fig:radial1} predicts.
This raises the possibility that the overall fraction of first stars may
even be higher in the halo than in the bulge!

These results have a number of immediate implications.  First,
as stars enriched only by primordial stars (which trace the overall distribution
of both first and second stars in our study) are found throughout
the Galactic halo, measured abundances of
metal-poor stars in the Galactic halo
should be taken as directly constraining the properties of primordial
objects.   
Furthermore since only a slightly larger fraction of stars in the
Galactic bulge are primordial, there is no compelling theoretical
reason that observations should have to focus in this extremely
difficult region.   Instead the lack of metal-free stars observed in
the halo is likely to imply a real lower mass limit in the metal free
IMF of at least $0.8 \msun,$ the mass of a low-metallicity star  with
a lifetime comparable to the Hubble time (Fagotto \etal 1994).
In fact the $\sim 10^{-6} M_\odot$ pc$^{-3}$ value for PopIII
stars in the solar neighborhood we compute is so high as compared to the 
observed stellar mass density of $5 \times 10^{-5} M_\odot$ pc$^{-3}$
(Preston, Shectman, \& Beers 1991) that several primordial
$0.8 \msun$ stars would have been observed even if $f_*^{III}$ were
over an order of magnitude lower than the $0.1$ value we
used to normalize our approach.
Finally, it means that there is a good chance that many of the stars
with extremely unusual abundances, such as HE0107-5240 (Christlieb et
al.\ 2002) and HE1327-2326  (Frebel et al.\ 2005; Aoki et al.\ 2006),
may in fact represent the products of self-enrichment in primordial
star clusters, which display the yields of a single neighboring PopIII
star (\eg Umeda \& Nomoto 2003; Suda et al.\ 2004; Iwamoto et
al. 2005; Tsujimoto \& Shigeyama 2006).
Note that this widespread PopIII star distribution is also consistent
with studies of the lithium-7 abundance of metal-poor [Fe/H] $<$ -1.3
stars, which suggest that a large fraction of the baryonic
matter in the early halo may have been processed through
such stars (Piau \etal 2006).

\begin{figure}
\plotone{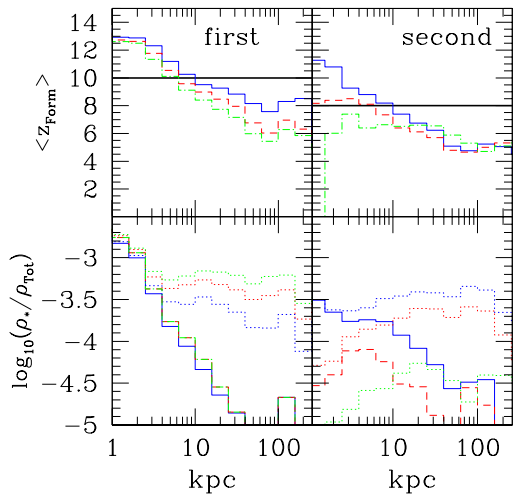}
\caption{{\em Top:} Average formation redshift of first
and second stars in our Lagrangian models.  As in Fig. \ref{fig:radial1},
models L2, L3, and L4 are shown by the solid,
short-dashed, and dot-dashed lines respectively.   {\em Bottom:}
Normalized number density of 
first stars forming above $z=10$ and second
stars forming above $z=8$, with lines as in the upper panels.
For comparison these plots also include the total mass fraction
of first and second stars, given by the dotted lines.}
\label{fig:z1}
\end{figure}

In the upper left panel of Figure \ref{fig:z1} we plot the mean
redshift of formation of the first stars as a function of radius.
While this is not a directly observable quantity, it nevertheless
helps in contrasting our results with previous estimates that assume a
single formation redshift or $\sigma$ for the  first stars.  In all
models $\left<z_{\rm Form} \right>$ decreases monotonically with
radius, moving from  $\left<z_{\rm Form} \right> \approx 13$ near the
Galactic center, to $\left<z_{\rm Form} \right> < 8$ at a distance of
100 kpc.  In the lower left panel of this figure, we plot the
normalized radial mass density of first stars with formation  redshifts
above 10.  This shows that essentially all of the earliest forming
first stars are located near the Galactic center, and that their
relative number density drops by at least a factor of 30 as we move
towards 8 kpc.

These results are reminiscent of those of White \& Springel (2000) and
Diemand, Madau, \& Moore (2005), who used N-body simulations to show
that the oldest stars are very strongly concentrated near the Galactic
center (see also Miralda-Escud\' e 2000).  Interestingly, White \&
Springel (2000) also estimated the history of very metal-poor Galactic
stars  by rescaling an N-body cluster simulation to Milky Way scales,
and associating PopIII stars with {\em all} particles collapsing into
$T_{\rm vir} = 10^4$K halos.   They found that  these stars had  a
mean formation redshift of $z \approx 4$ and a spatial distribution
closely following the overall dark matter distribution.  Our more
detailed modeling shows that metal free star formation displays
elements of both these scenarios.  Primordial stars form in the smallest
halos that can cool efficiently, but only within a subset of these
halos that is biased to the highest redshifts. These stars do not follow 
the overall mass distribution, but display a only weak radial density
gradient. And their overall formation redshifts, while always high,
nevertheless drop off strongly as a function of radius.

Finally, the upper right panel of Figure \ref{fig:z1} shows the
average formation redshift of second stars in our simulation.    Again
we remind the reader that these are the stars that contain metal
from primordial stars but are not formed in the same star clusters.
These stars display a similar radial trend as the first stars, but
shifted to somewhat lower formation redshifts.  In this case, the
majority of these stars in the solar neighborhood are formed at
redshifts below 8, as shown in the lower right panel of this figure.

\subsection{Model Tests and Refinements}

\begin{figure}
\plotone{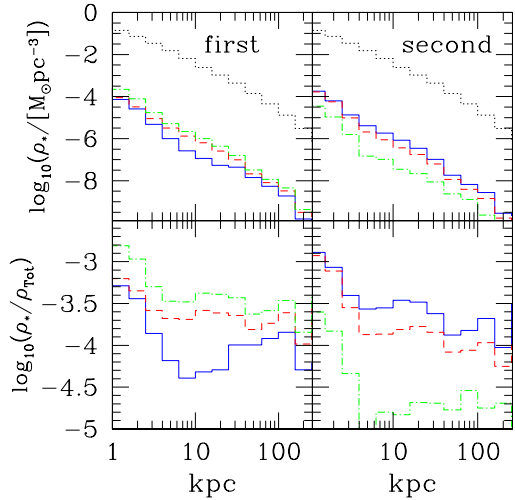}
\caption{Radial profile of  first and second stars at $z=0$ in 
Eulerian models 2E (solid), 3E (dashed), and 4E (dot-dashed).
Panels are as in
Fig \ref{fig:radial1}.  Compared to the Lagrangian models, the
Eulerian models show a somewhat smaller number of first stars and a
more strongly  centrally-peaked concentration of second stars.}
\label{fig:radial2}
\end{figure}

As our results in  Figs.\ \ref{fig:Dist1} and \ref{fig:radial1} have a
number of wide-reaching implications, it is important to understand to
what degree they are dependent on our assumptions.  The primary
simplification in our model is our use of a Lagrangian coordinate
system combined with a bubble model that assumes expansion into the
Hubble flow.  To asses this approximation, we show in Figure
\ref{fig:radial2} the results of a calculation that replaces the
initial coordinates with ``Eulerian positions'' that re-center each
bubble from output to output, as described in \S2.2.  As this approach
combines the large bubble radii computed in the absence of infall,
with the small separation computed by accounting for
the proper motions of halos, it can be taken as providing a lower
limit on the distribution of first stars.

Thus, it is not surprising that the densities of first
stars in these models are lower than their Lagrangian counterparts.
Yet, in no cases are these differences dramatic.   Instead, the radial
distributions in the $\E3 = 10^{-3}$ (model E3 in Table 1) and $10^{-4}$ (model E4) 
cases  in Figure \ref{fig:radial1} show only a slightly lower overall  normalization
and a marginally steeper radial trend than seen in the Lagrangian case.
The $\E3 = 10^{-2}$ model also shows an overall radial trend that is
very similar to that in Fig.\ \ref{fig:radial1}, although it contains
a gap from 5-20 kpc with a factor of $\approx 2$ deficit in stars.
Thus, even in this most extreme of any of our models, the Galactic
halo contains a substantial number of the remnants of PopIII objects.

On the other hand,  much larger differences between our Eulerian and
Lagrangian models are visible in the distribution of second stars.
These stars are substantially more  centrally-concentrated in the
Eulerian models, and in the $\E3 = 10^{-4}$ case this difference is so
severe that very few second stars are located in the Galactic halo.
This represents the largest uncertainty in our approach, and thus  for
the limited case of a weak-feedback model, we are unable to draw any
clear conclusions as to the halo distribution of second stars.

\begin{figure}
\plotone{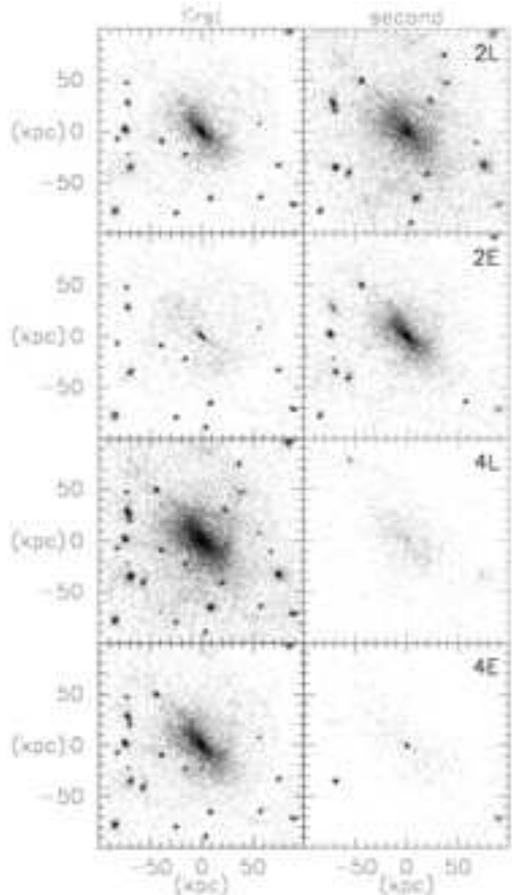}
\caption{Comparison of the $z=0$ distribution of first and second
stars in our Lagrangian and Eulerian models.  The top two rows compare
the results of strong feedback models ($\E3 = 10^{-2}$) with a
Lagrangian distance (top row) and an Eulerian distance (second row).
The bottom two rows show a similar comparison for weak feedback models
($\E3 = 10^{-4}$) with Lagrangian (third row) and Eulerian (bottom
row) distances.}
\label{fig:Dist2}
\end{figure}

The differences between the Lagrangian and Eulerian approaches are
further investigated in Figure \ref{fig:Dist2}, which shows the
spatial distribution of first and second stars in both models.   Here
we see the same overall trends as in the radial profiles.  In the
strong feedback case, first and second stars are more concentrated
than in the Eulerian approach, although  in both models significant
numbers of these stars are found in the halo and in satellite
galaxies.  In the weak feedback case, on the other hand, the
differences in the first-star distribution are minor, but the lack of
second stars in the  Galactic halo in the Eulerian model is considerable.
Finally, apart from the difference in radial trends, the detailed
distribution of satellites containing PopIII stars differ somewhat
between each of these approaches.  Although in general, there are
fewer PopIII satellites in the Eulerian cases, there are also
occasionally PopIII satellites in these models without Lagrangian
counterparts.

\begin{figure}
\plotone{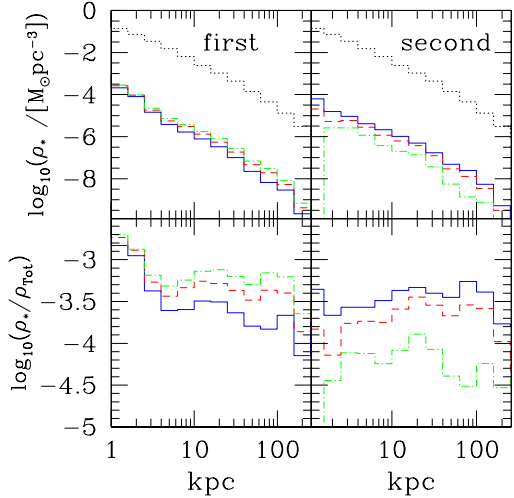}
\caption{Radial profile of first and second stars at $z=0$ in our
Lagrangian models, but now only associating stars with the mass in the
central 10\% of each halo.   Panels are as in Figs.\
\ref{fig:radial1} 
and \ref{fig:radial2}, while the lines represent
models 2Lc (solid), 3Lc (dashed), and 4Lc (dot-dashed).
In all cases, the radial
profiles are consistent with those in our fiducial  approach, shown
in Fig.\ \ref{fig:radial1}.}
\label{fig:radial3}
\end{figure}

\begin{figure}
\plotone{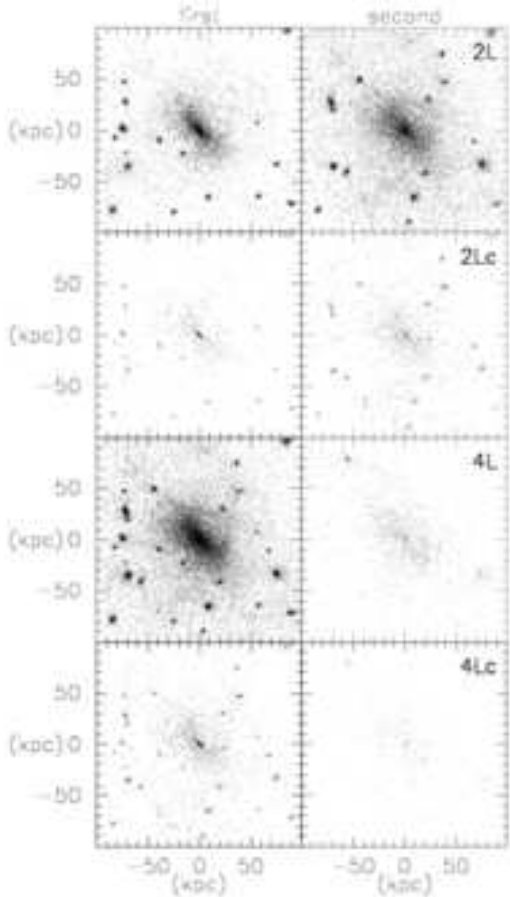}
\caption{Comparison of the $z=0$ distribution of first and second
stars between our fiducial models and Lagrangian models in which stars
are associated with the mass in the inner 10\% of each halo.
Consistent with Fig.\ \ref{fig:radial3}, there are no strong
difference between these two  approaches.}
\label{fig:Dist3}
\end{figure}

A second important issue is our association of stars with the overall
distribution of dark matter in each halo.   In reality, the gas in a
given object will condense into a rotationally-supported disk whose
radius is substantially smaller that its dark matter virial radius.
Furthermore, due to the increased impact of tidal effects on the most
loosely bound particles, it is possible that stars forming in such
disks may end up distributed significantly differently at $z=0$ than 
their associated dark matter.

To account for this possibility, we repeated our Lagrangian model, but
now only associating stars with the inner 10\% of the dark matter
particles in first and second-generation objects, while leaving
the number of stars formed in each halo fixed.  The resulting
radial distributions and spatial profiles are shown in Figures
\ref{fig:radial3} and \ref{fig:Dist3}, respectively.  Here we see that
while limiting the number of particles associated with stars leads to
larger statistical fluctuations, the distribution of first and second
stars otherwise remains the same as in our fiducial modeling.
Carrying out the same exercise in the Eulerian case yields similar
agreement, and thus our results do not seem to be affected by
differences in tides across halos.

\begin{figure}
\plotone{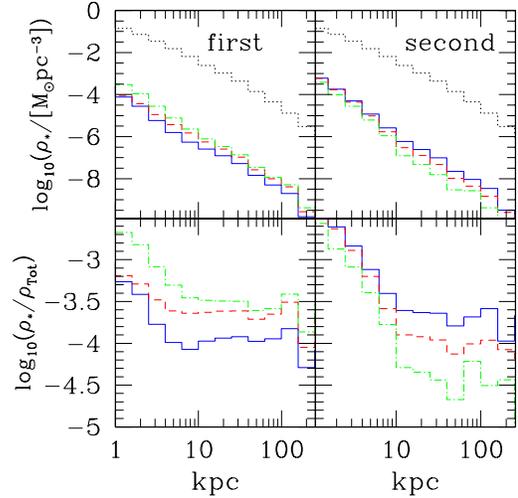}
\caption{Radial profile of first and second stars at $z=0$ in 
Lagrangian models with a $0.15$ choice
of linking length.  Panels are as in figure 
\ref{fig:radial1} 
and \ref{fig:radial2}, and the lines represent
models 2L15 (solid), 3L15 (dashed), and 4L15 (dot-dashed).
For the first stars, the radial
profiles are consistent with those in our fiducial, $b=0.2$  
approach, shown in Fig.\ \ref{fig:radial1}.  However, choosing
$b=0.15$ results in more second-stars being identified within
$\approx 8$ kpc, which is primarily 
due to a large high-redshift object that is split into several 
sub-halos with this smaller choice of linking length. }
\label{fig:radial4}
\end{figure}

Another important parameter in our analysis is our choice
of $b = 0.2$ linking length when identifying halos.  To test the
dependence of our results on this choice, we generated new
lists using a much smaller $0.15$ length, which associates
halos with objects almost twice as dense as in the standard
picture.   In Figure \ref{fig:radial4}  we show the results of this
calculation in the Lagrangian case (models 2L15-4L15).  

\begin{figure}
\plotone{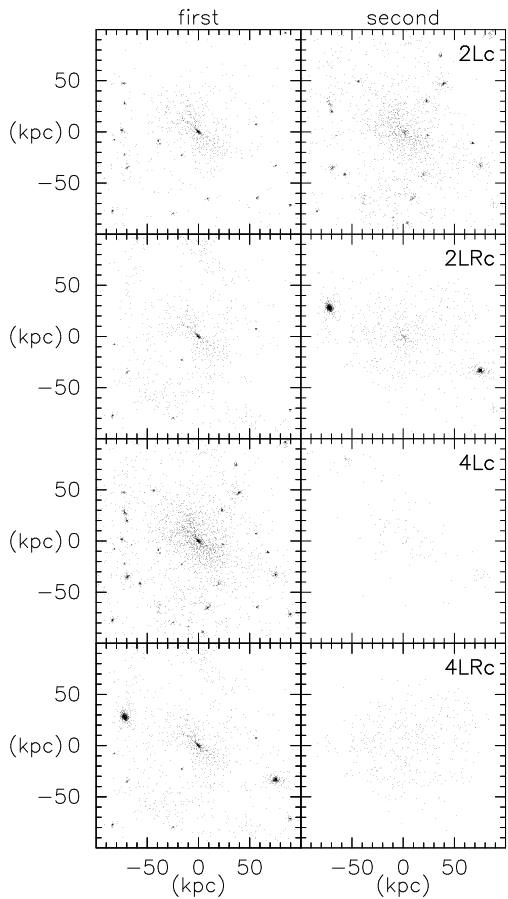}
\caption{Comparison between the spatial distribution of stars in
models with a fixed minimum virial temperature
(2Lc and 4Lc), and models with a gradually increasing
threshold associated with reionization (2LRc and 4LRc).
While incorporating reionization leads to a far smaller number of
first and second satellites, the overall distribution of such stars
remains widespread.}
\label{fig:plotreion}
\end{figure}

\begin{figure}
\plotone{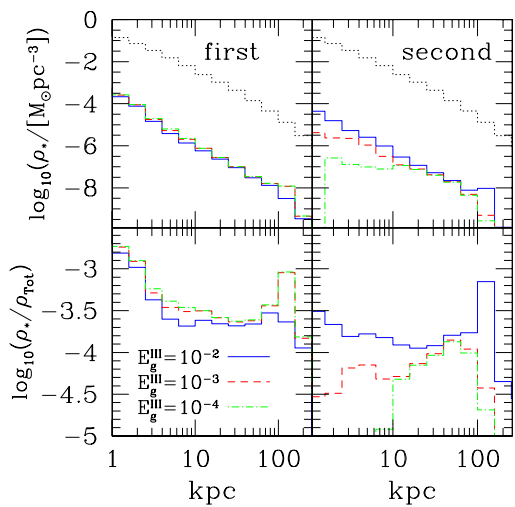}
\caption{Radial profile of first and second stars at $z=0$ in 
Lagrangian models that include reionization.
Panels are as in Figures 
\ref{fig:radial1} 
and \ref{fig:radial2}, and the lines represent
models 2LRc (solid), 3LRc (dashed), and 4LRc (dot-dashed).
Increasing the filtering scale from $10^4$K at $z=9$ to $5 \times 10^5$K at $z=7$ 
has only a weak effect on our results.}
\label{fig:radialreion}
\end{figure}

\begin{figure}
\plotone{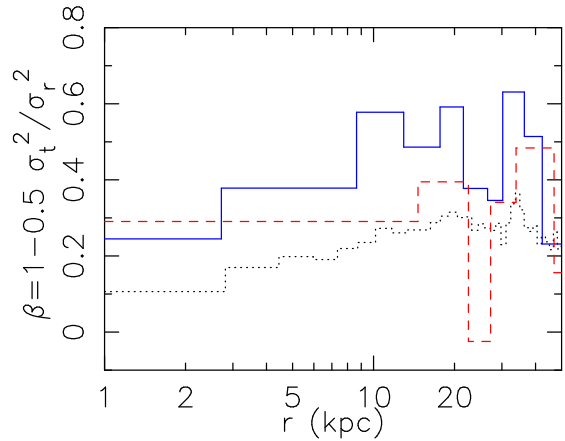}
\caption{The anisotropy parameter of the velocity dispersion,
$\beta$,  as a function of radius,  for all the particles (dotted),
first star  particles (solid), and second star particles (dashed).}
\label{fig:beta}
\end{figure}

\begin{figure*}
\epsscale{0.8}
\plotone{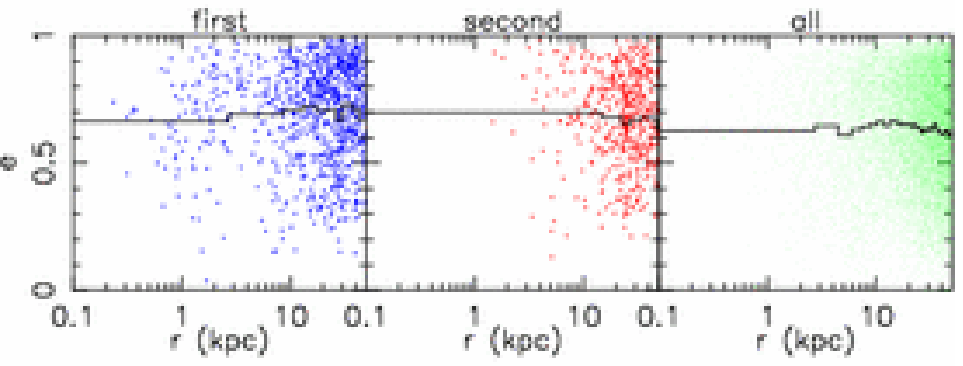}
\caption{The eccentricity,  $e=(r_{\rm max}-r_{\rm min})/(r_{\rm
max}+r_{\rm min})$, as a function of the radius at $z=0$ for first
(left panel) and second (center panel) star praticles, contrasted with
the full distribution (right panel).   In all cases the line gives the
average value as a function of radius.}
\label{fig:ecc}
\end{figure*}

From the left panels of this figure, we see that while adopting
this more stringent criteria reduces the number of PopIII
stars somewhat, this effect is no stronger than a factor
of $\approx 2.$  Furthermore, the overall radial profiles
in the $b=0.15$ case are very similar, if not flatter than
in the fiducial model, indicating that the choice of linking
length has no impact on our conclusions about the radial
distribution of first stars.   On the other hand,
changing $b$ results in a substantial increase the number
of second-stars within $8$ kpc as shown in the right panels.   
This can be traced to the presence of 
a large object at $z=11$, which was
identified as a single large halo in the $b=0.2$ case, but
split into several smaller objects in the $b=0.15$ case.   
While in the fiducial case, this large object was a direct
descendant of a first stars halo, and thus excluded from
the second-stars count,  in the $0.15$ case, most of
the mass in this object was tagged as neighboring halos,
and was added to the second-stars count.   While this difference
is consistent with our approach to counting 
first and second stars, it nevertheless illustrates some of
the ambiguities involved in distinguishing stars that are formed in
gas initially enriched by primordial stars, from stars that are
formed within purely self-enriched primordial gas.
Note however, that
at Galactocentric radii $\gtrsim 8$ kpc, the overall distributions
of second stars are similar between the models with different
choices of linking length.

Another concern is the possibility that our models are 
assigning first and second stars to {\em all} halos that
collapse in primordial areas or areas enriched only
by primordial stars.  However, a one-to-one association
between dark matter halos and Milky-Way 
satellites leads to a large excess with respect
to their observed numbers, which is commonly referred to
as the missing satellite problem (\eg Moore \etal 1999).
Although ram pressure stripping before star formation (\eg
Scannapieco, Ferrara, \& Broadhurst 2002)
tidal stripping after star formation (\eg Kravstov, Gnedin, \& Klypin 2004)
and other issues are important in resolving this issue, 
the largest source of suppression of dwarf galaxy formation is
the increase in IGM thermal pressure following reionization
(\eg. Bullock, Kravstov,\& Weinberg 2000).   While reionization
will clearly have the largest effect on the very latest
forming stars, rather that the objects in which we are most 
concerned here, the relatively large number of satellites
seen in plots such as Figs.\ \ref{fig:anim} and \ref{fig:Dist1}
indicates that our results may nevertheless suffer from such
a excess of Galactic satellites.

Recently Benson \etal (2002), have shown that incorporating 
the increase in IGM pressure associated with reionization
into semi-analytical models of galaxy formation 
helps to bring their faint end galaxy counts into rough agreement with observations.
Following their lead then, we carried out a set of reionization runs, in 
which we imposed the minimum filtering mass which approximates that
computed in Gnedin (2000).
In this case the minimum virial temperature in our calculation is increased
linearly from the $T_{\rm vir} \geq  10^4$ K molecular hydrogen limit at $z=9$, to 
$T_{\rm vir} \geq 5 \times 10^{4}$ below the final overlap redshift of $z=7$.
The resulting spatial distributions and radial profiles are shown in 
Figures \ref{fig:plotreion} and \ref{fig:radialreion} respectively.

From these plots, it is clear that incorporating such suppression leads to far fewer 
first and second satellites, moving from  $\approx 25$ such objects with 100 kpc
in our fiducial models, to less than 10 in the reionization models.  Nevertheless,
as shown in Figure \ref{fig:radialreion} the overall radial profile
of first and second stars are quite similar to our fiducial results.  With
metal-free stars (first stars), and stars that contain the products of metal-free 
stars (first + second stars) forming at all Galactocentric radii.
Thus we do not expect the details of suppression of dwarf galaxy formation
to have a strong effect on our conclusions.

Finally, we note that the cosmological parameters used in our model
have recently been revised by the 3-year data release from the {\em
Wilkinson Microwave Anisotropy Probe} experiment (Spergel \etal 2006).
In general, we expect that the low $\sigma_8$ and $n$ values implied
by these measurements should strenghten our conclusions. Indeed, the
reduced power on  small scales leads to a delay in the onset of star
formation and metal  enrichment in our $T_{\rm vir} > 10^4$ K halos. As a
consequence, the first stars will form at more recent epochs than in our
fiducial model, and are likely to give a larger contribution to the
halo population.

\subsection{Kinematics of First and Second Stars}

Our simulation results offer not only the spatial distribution of
first and second stars, but also their kinematics.  In this section,
we present the results of a representative Lagrangian model, model
2Lc.  Figure \ref{fig:beta} demonstrates the predicted radial profile
of  the anisotropy parameter, $\beta$, of the velocity dispersions
first and second star particles as compared to the overall
distribution.  Here, $\beta=1-0.5 \sigma_t^2/\sigma_r^2$, where
$\sigma_r$ is the radial velocity dispersion and
$\sigma_t^2=\sigma_{\theta}^2+\sigma_{\phi}^2$ is  the tangential
velocity dispersion.  The bin sizes are adjusted so that each consists
of 100 particles for first and second star case and 1000 particles for
the overall distribution. Here, we focus on the results within 50
kpc. Overall DM particles show almost isotropic velocity dispersion in
the central region, and $\beta$ gradually increases, i.e., the radial
velocity dispersion becomes more dominant, with increasing
radius. This is consistent with previous equally high-resolution
N-body simulations of a Milky Way size halo (e.g., Diemand et al.\
2005).  Compared with the overall particles, the first star particles
have higher $\beta$, i.e., the radial velocity dispersion is more
dominant, especially at smaller radii.  This is expected, because the
first star particles  in the inner region form in halos closer to the
density peak  of the main system at high redshifts,  and fall into the
main system at earlier epochs.  Second stars have a degree of
anisotropy between the first particles and the overall distribution.

We also measure the eccentricity of the orbits of the first and second
stars in Figure \ref{fig:ecc}.  To compute these values, we run an
N-body simulation for these particles in the fixed potential of the
final system. Here, we fix the position of all the particles within
the virial radius, and follow the evolution of the first and second
particles from their position and velocity at $z=0$.  We then compute
the motion of each particle over a full orbit after its first passing
through pericenter or apocenter, and measure the maximum, $r_{\rm
max}$, and minimum, $r_{\rm min}$, radii over this full trajectory.
The eccentricity is then defined as, $e=(r_{\rm max}-r_{\rm
min})/(r_{\rm max}+r_{\rm min})$.   Figure \ref{fig:ecc} shows that
both first and second particles have a  relatively high eccentricity
($\left< e \right>  \approx 0.7$). This slightly exceeds the mean
value of the overall distribution, which is roughly  consistent with
the observed value of $\left<e \right> \approx 0.6$ (Chiba \& Beers
2000).

\section{Conclusions}

Theoretical models of PopIII star formation have focused on their
overall cosmological distribution, or their contribution to the global
star formation history of the Galaxy.  But $z=0$ observations are not
of this  type.  As PopIII stars would have been carrying out nuclear
burning for roughly a Hubble time, they must be faint, low-mass stars
that are only observable nearby, and in environments without
significant dust extinction or crowding from higher-mass stars.  For
this reason searches for metal-poor stars have been targeted  only in
the halo of the Milky Way, with an emphasis on halo stars roughly in
the solar neighborhood.     This means that comparisons  between
theory and observations have  been dependent on extremely uncertain
extrapolations, drawn from measurements in a limited environment.

In this work, we have carried out the first theoretical investigation
able to quantify the impact of this extrapolation.    Our method
relies upon two key ingredients.  The first of these is an extremely
high-resolution N-body simulation of the Milky Way, which is capable
of reliably capturing the formation of objects down to the $10^4$ K
atomic cooling limit.  The second is a simple and flexible analytic
model of metal-dispersal, which allows us to quickly identify regions
of metal-free star formation, despite the enormous uncertainties in
the properties of the first stars and supernovae.

Our results are robust and optimistic.  Despite changing the
efficiency of metal dispersal by two orders of magnitude, adopting
widely different estimates for the distances between neighboring
objects, and varying our assumptions as to the regions of
star-formation within collapsed objects, local measurements place
substantial constraints on PopIII star formation in all of our models.
In particular we have found that:

(1) Galactic PopIII star formation occurs over a range of redshifts,
as metal pollution takes a substantial amount of time to affect all
the  Milky Way progenitors.  While Galactic PopIII star formation  is
likely to have peaked at  $z \approx 10$, it should have continued  at
appreciable values down to  $z \approx 5$, paralleling the overall
cosmological evolution.

(2) The mass fraction of PopIII stars in the Galactic halo near the
solar orbit is only slightly lower than it is in the bulge.  Thus, if
they  have sufficiently long lifetimes, a significant number of stars
formed in  initially primordial star clusters should be found in the
Galactic halo.  This means that there is no compelling theoretical
reason to motivate  observational searches in more difficult
environments, and present observations should be taken as directly
constraining the properties of PopIII stars.  This picture is also
consistent  with observational studies of the lithium-7 abundance of
metal-poor stars, which suggests that  a large fraction of the
baryonic matter in the early halo  may have been proceed by PopIII
stars.

(3) While our models are not able to quantify self-enrichment within
individual star clusters in detail, we expect that stars enriched
purely by metal-free stars in the same cluster will exist cospatially 
with the ``first stars'' in our model.  Furthermore we are  able to make
statements about the distribution of ``second stars'' that do not form
cospatially with PopIII stars, but rather form out of gas that  is
enriched by neighboring clusters of PopIII stars.  In almost all of
our models,  the mass fraction of such stars in the Galactic halo is
comparable to that of PopIII stars, although there are very few second
stars in our weakest $\E3 = 10^{-4}$, Eulerian case.   Unless the
efficiency of PopIII metal ejection was at the lower end of the
allowed range, many of these stars should also be found nearby.

(4) We find that the final distribution of first and second stars remains 
practically unchanged when we move from our fiducial model, which
associates stars with all dark-matter particles in each first and second
generation object, to a centrally-concentrated model, which associates
stars with only the inner 10\% of particles in each object. 
Thus we do not expect the spatial distribution of stars within each 
first and second generation object to affect their distribution today.

(5) Incorporating the effects of reionization leads to
an overall smaller number of satellite galaxies containing first and
second stars, and helps to address the ``missing satellite problem.''  However
including such effects does not change the overall widespread distribution of 
these stars.

While stars are the sources of all metals, metals need not be
contained in every star.  Furthermore, we have shown here that stars
without metals  need not to be located near the Galactic center.  Rather
observations of metal-poor halo stars are providing us with
direct constraints on the properties of the first stars.  Though
primordial stars were formed in the distant past, some of our greatest clues
to the process of primordial star formation are likely to come from
our local Galactic neighborhood.

\acknowledgements

We thank Tim Beers, Norbert Christlieb, Anna Frebel, 
Falk Herwig, J. P. Ostiker,
Eric Pfahl, Sean Ryan, and Kim Venn for helpful comments that 
significantly improved this work, and we are particularly thankful
to Ken Freeman, who suggested the kinematic tests presented in \S 3.3.
We also with to thank the referee for a careful reading of this manuscript,
which resulted in several important improvements.
We acknowledge the Astronomical Data Analysis Center of
the National Astronomical Observatory, Japan (project ID: wmn14a), the
Institute of Space and Astronautical Science of Japan Aerospace
Exploration Agency,  and the Australian and Victorian Partnerships for
Advanced Computing, where the numerical computations for this paper
were performed. ES was supported by the National Science Foundation
under grant PHY99-07949.  DK thanks the JSPS for financial support
through a Postdoctoral Fellowship for research abroad.

\fontsize{10}{10pt}\selectfont

\end{document}